\documentclass{article}
\usepackage{amsmath,amsthm,amsfonts,amscd,eucal}

\usepackage{amssymb,amsmath,amsfonts,amscd,eucal}
\usepackage{graphicx}

\numberwithin{equation}{section}

\def\Var{{\rm Var}}

\def\Cov{{\rm Cov}}

\newtheorem{Thm}{Theorem}[section]

\newtheorem{Prop}[Thm]{Proposition}
\newtheorem{Lemma}[Thm]{Lemma}

\theoremstyle{definition}
\newtheorem{Dfn}[Thm]{Definition}

\theoremstyle{remark}
\newtheorem{rem}[Thm]{Remark} 
\newtheorem{ack}{Acknowledgement} 
\begin{document}

\title{Uncertainty Principle and Quantum Fisher Information}

\author{Paolo Gibilisco$^{1}$, Tommaso Isola$^{2}$\\
$^{1}$Dipartimento SEFEMEQ and Centro V.Volterra,\\
Facolt\`a di Economia,\\ Universit\`a di Roma ``Tor Vergata", Via
Columbia 2, 00133 Rome, Italy. \\
$^{2}$Dipartimento di Matematica,\\ Universit\`a di Roma ``Tor
Vergata'', I--00133 Roma, Italy.}

\maketitle

\begin{abstract}
    A family of inequalities, related to the uncertainty principle,
    has been recently proved by S. Luo, Z. Zhang, Q. Zhang, H. Kosaki,
    K. Yanagi, S. Furuichi and K. Kuriyama.  We show that the
    inequalities have a geometric interpretation in terms of quantum
    Fisher information.  Using this formulation one may naturally ask
    if this family of inequalities can be further extendend, for
    example to the $RLD$ quantum Fisher information.  We show that
    this is impossible by producing a family of counterexamples.
\end{abstract}


\section{Introduction}



Noncommutativity in quantum probability has far-reaching consequences. 
One of the most important is the Heisenberg uncertainty principle
$$
{\rm Var}_{\rho}(A)\cdot{\rm Var}_{\rho}(B)\geq \frac{1}{4}\vert {\rm
Tr}(\rho[A,B])\vert^2 .
$$
No such lower bound for the variance of pairs of random variables
exists in classical probability.  Schr{\"o}dinger proved a stronger
inequality involving covariance
$$
{\rm Var}_{\rho}(A)\cdot{\rm Var}_{\rho}(B)-|{\rm Re}\{{\rm Cov}_{\rho}(A,B) \}|^2 \geq
\frac{1}{4}\vert {\rm Tr}(\rho[A,B])\vert^2 .
$$

Recently S. Luo and Q. Zhang proved a different kind of uncertainty
principle (see Luo and Q.Zhang (2004), Theorem 2), in the
Schr{\"o}dinger form, where the lower bound appears because the
variables $A,B$ do not commute with the state $\rho$ (in contrast with
the standard uncertainty principle where the bound depends on the
commutator $[A,B]$).

The inequality was conjectured by S. Luo himself and Z. Zhang in a
previous paper (Luo and Z.Zhang (2004)).  These authors suggest there
that ``the result may be interpreted as a quantification of certain
aspect of the Wigner-Araki-Yanase theorem for quantum measurement,
which states that observables not commuting with a conserved quantity
cannot be measured exactly" (see Wigner (1952), Araki and Yanase
(1960), Ozawa (2002)).  The inequality has been recently generalized
in Kosaki (2005) and Yanagi-Furuichi-Kuriyama (2005).  The final
result is
$$
{\rm Var}_{\rho}(A)\cdot{\rm Var}_{\rho}(B)-|{\rm Re}\{{\rm
Cov}_{\rho}(A,B) \}|^2\geq I_{\rho,\beta}(A)I_{\rho,\beta}(B)-|{\rm
Re}\{{\rm Corr}_{\rho, \beta}(A,B) \}|^2
$$
where $I$ and Corr are given by the Wigner-Yanase-Dyson skew
information (see Section \ref{Section 3} below).

The purpose of this paper is to put the above inequality in a more
geometric form by means of quantum Fisher information (namely the
monotone metrics classified by Petz).  In this way the lower bound
will appear as a simple function of the area spanned by the
commutators $i[A,\rho], i[B,\rho]$ in the tangent space to the state
$\rho$, provided the state space is equipped with a suitable monotone
metric (see Theorem \ref{MR}).  At this point it is natural to ask
whether such an inequality holds for other quantum Fisher informations
in the Wigner-Yanase-Dyson class (like the $RLD$-metric for example). 
The answer turns out to be negative and a general counterexample is
given in Proposition \ref{CE}.

In the final section we discuss some open problems related to the
subject.

\section{Schr{\"o}dinger and Heisenberg Uncertainty Principles}

Let $M_n:=M_n(\mathbb{C})$ (resp.$M_{n,sa}:=M_n(\mathbb{C})_{sa}$) be
the set of all $n \times n$ complex matrices (resp.  all $n \times n$
self-adjoint matrices).  We shall denote general matrices by $X,Y,...$
while letters $A,B,...$ will be used for self-adjoint matrices.  Let
${\cal D}_n$ be the set of strictly positive elements of $M_n$ while
${\cal D}_n^1 \subset {\cal D}_n$ is the set of strictly positive
density matrices namely
$$
{\cal D}_n^1=\{\rho \in M_n \vert {\rm Tr} \rho=1, \, \rho>0 \}.
$$

\begin{Prop}
The correspondence
$$
M_n \times M_n \ni (X,Y) \to \langle X,Y \rangle := {\rm Tr}(\rho
XY^*)-{\rm Tr}(\rho X)\cdot \overline{{\rm Tr}(\rho Y)}
$$
is a positive sesquilinear form.
\end{Prop}

\noindent
As usual commutators and anticommutators are defined as $[X,Y]=XY-YX$ , $\{X,Y\}=XY+YX$.

\begin{Dfn}
Suppose that $\rho \in {\cal D}_n^1 $ is fixed. Define $X_0:=X-{\rm Tr}(\rho X)I$.
\end{Dfn}

\begin{Dfn}
For $A,B \in M_{n,sa}$ and $\rho \in {\cal D}_n^1$ define covariance and variance as
$$
{\rm Cov}_{\rho}(A,B):=\langle A,B \rangle={\rm Tr}(\rho A B)-{\rm
Tr}(\rho A)\cdot{\rm Tr}(\rho B)={\rm Tr}(\rho A_0 B_0)
$$
$$
{\rm Var}_{\rho}(A):=\langle A,A \rangle={\rm Tr}(\rho A^2)-{\rm
Tr}(\rho A)^2={\rm Tr}(\rho A_0^2) .
$$
\end{Dfn}

Note that for $A,B \in M_{n,sa}$ and $\rho \in {\cal D}_n^1$ one has
$$
{\rm Re}({\rm Tr}(\rho A B))=\frac{1}{2}{\rm Tr}(\rho \{A,B \})
\qquad \qquad
{\rm Im}({\rm Tr}(\rho A B))=\frac{1}{2i}{\rm Tr}(\rho[A,B]).
$$
Since ${\rm Cov}_{\rho}(A,B)= \overline{{\rm Cov}_{\rho}(B,A)}$ then
$$
2{{\mathop{\rm Re}\nolimits} \left\{ {\hbox{Cov}_\rho \left( {A,B}
\right)} \right\}}= {\rm Cov}_{\rho}(A,B)+{\rm Cov}_{\rho}(B,A) .
$$
As a consequence of Cauchy-Schwartz inequality one can derive the
Schr{\"o}dinger and Heisenberg Uncertainty Principles that are given
in the following 

\begin{Thm} (see Schrodinger (1930)) For $A,B \in
M_{n,sa}$ and $\rho \in {\cal D}_n^1$ one has
$$
{\rm Var}_{\rho}(A)\cdot{\rm Var}_{\rho}(B)- |{\rm Re}\{{\rm Cov}_{\rho}(A,B) \}|^2\geq
\frac{1}{4}\vert {\rm Tr}(\rho[A,B])\vert^2
$$
that implies
$$
{\rm Var}_{\rho}(A)\cdot{\rm Var}_{\rho}(B)\geq \frac{1}{4}\vert {\rm Tr}(\rho[A,B])\vert^2 .
$$
\end{Thm}

\begin{Dfn}
Set
$$
{\cal S}_{\rho}(A,B):={\rm Var}_{\rho}(A)\cdot{\rm
Var}_{\rho}(B)-|{\rm Re}\{{\rm Cov}_{\rho}(A,B) \}|^2 .
$$
\end{Dfn}

\begin{rem}
With the above definition the Schr{\"o}dinger Uncertainty Principle takes the form
$$
{\cal S}_{\rho}(A,B) \geq
\frac{1}{4}\vert {\rm Tr}(\rho[A,B])\vert^2 .
$$
\end{rem}

Let us try to see this situation in general.

\begin{Dfn}
Let $F:{\cal D}_n^1 \times M_{n,sa} \times M_{n,sa} \to \mathbb{R}$ be
a function (denoted as $F_{\rho}(A,B)$) such that
$$
{\cal S}_{\rho}(A,B) \geq F_{\rho}(A,B).
$$
Then we say that $F$ is an Uncertainty Principle Function (shortly UPF).
\end{Dfn}

\noindent
Problem: are there nontrivial UPF different from $\frac{1}{4}\vert {\rm Tr}(\rho[A,B])\vert^2$?

\noindent More specifically: $[A,B] \not=0$ gives a non-trivial bound
for ${\cal S}_{\rho}(A,B)$.  Does the condition $[A,\rho], [B,\rho]
\not=0$ give a non-trivial bound for ${\cal S}_{\rho}(A,B)$?

\section{An inequality related to uncertainty principle} \label{Section 3}

\begin{Dfn}
For $A,B \in M_{n,sa}$, $\rho \in {\cal D}_n^1$ and $\beta \in (0,1)$ set
$$
{\rm Corr}_{\rho,\beta}(A,B):={\rm Tr}(\rho AB)-{\rm Tr}(\rho^{\beta}A\rho^{1-\beta}B).
$$
\end{Dfn}

\begin{Dfn}
The Wigner-Yanase-Dyson information is defined as
$$
I_{\rho,\beta}(A):={\rm Corr}_{\rho,\beta}(A,A)=-\frac{1}{2}{\rm
Tr}([\rho^{\beta},A]\cdot[\rho^{1-\beta},A]).
$$
\end{Dfn}

With direct calculation one can prove the following

\begin{Lemma} \label{Corr}
$$
2{\rm Re} \{ {\rm Corr}_{\rho,\beta} ( A,B) \} ={\rm
Corr}_{\rho,\beta}(A,B)+{\rm Corr}_{\rho,\beta}(B,A) =-{\rm
Tr}([\rho^{\beta},A]\cdot[\rho^{1-\beta},B]),
$$
$$
{\rm Corr}_{\rho,\beta} \left( {A,B} \right)={\rm Cov}_{\rho} \left(
{A,B} \right)-{\rm Tr}(\rho^{\beta}A_0\rho^{1-\beta}B_0).
$$
\end{Lemma}

\begin{Dfn}
$$
{\cal T}_{\rho,\beta}(A,B) :=I_{\rho,\beta}(A)I_{\rho,\beta}(B)-|{\rm
Re} \{ {\rm Corr}_{\rho,\beta} ( A,B ) \}|^2.
$$
\end{Dfn}

Note that ${\cal T}_{\rho,\beta}={\cal T}_{\rho,1-\beta}$ so one can
consider just $\beta \in (0,\frac{1}{2}]$.

In Luo and Q. Zhang (2005) the following result has been proved

\begin{Thm}\label{LZ} 
${\cal T}_{\rho,\frac{1}{2}}(A,B)$ is an UPF.
\end{Thm}

The theorem had been conjectured in Luo and Z. Zhang (2004).  A
generalization of Theorem \ref{LZ} has been given in Kosaki (2005) and
Yanagi {\sl et al.} (2005).

\begin{Thm} \label{K}
    ${\cal T}_{\rho,\beta}(A,B)$  is an UPF for any $\beta \in (0,\frac{1}{2}]$.
\end{Thm} 
\begin{proof}
 See Theorem \ref{order2}  in Section \ref{counterexample}
\end{proof}

 Kosaki proved Theorem \ref{K} by showing that ${\cal
 T}_{\rho,\beta}(A,B)$ is monotone increasing for $\beta \in
 (0,\frac{1}{2}]$.  Moreover he was able to prove that ${\cal
 S}_{\rho}(A,B)={\cal T}_{\rho,\beta}(A,B)$ iff $A_0, B_0$ are
 proportional.

In the next sections we try to give a more geometric form to Theorem \ref{K}.

\section{Quantum Fisher Informations}

 In the commutative case a Markov morphism is a stochastic map $T:
 \mathbb{R}^n \to \mathbb{R}^k$.  In the noncommutative case a Markov
 morphism is a completely positive and trace preserving operator $T:
 M_n \to M_k$.  Let
 $$
 {\cal P}_n := \{ \rho \in \mathbb{R}^n \vert \rho_i > 0 \} \qquad
 {\cal P}_n^1 :=\{ \rho \in \mathbb{R}^n \vert \sum \rho_i=1, \,
 \rho_i > 0 \}.
 $$

 In the commutative case a monotone metric is a family of riemannian
 metrics $g=\{g^n\}$ on $\{{\cal P}_n^1\}$, $n \in \mathbb{N}$, such that
 $$
 g^m_{T(\rho)}(TX,TX) \leq g^n_{\rho}(X,X)
 $$
 holds for every Markov morphism $T:\mathbb{R}^n \to \mathbb{R}^m$ and all $\rho
 \in {\cal P}_n^1$ and $X \in T_\rho {\cal P}_n^1$.

 In perfect analogy, a monotone metric in the noncommutative case is a
 family of riemannian metrics $g=\{g^n\}$ on $\{{\cal D}^1_n\}$, $n
 \in \mathbb{N}$, such that
 $$
 g^m_{T(\rho)}(TX,TX) \leq g^n_{\rho}(X,X)
 $$
 holds for every Markov morphism $T:M_n \to M_m$ and all $\rho \in
 {\cal D}^1_n$ and $X \in T_\rho {\cal D}^1_n$. 

 Let us recall that a function $f:(0,\infty)\to \mathbb{R}$ is said
 operator monotone if, for any $n\in N$, any $A$, $B\in M_n$ such
 that $0\leq A\leq B$, the inequalities $0\leq f(A)\leq f(B)$ hold. 
 An operator monotone function is said symmetric if $f(x):=xf(x^{-1})$. 
 With such operator monotone functions
 $f$ one associates the so-called Chentsov--Morotzova functions
 $$
 c_f(x,y):=\frac{1}{yf(xy^{-1})}\qquad\hbox{for}\qquad x,y>0.
 $$

 Define $L_{\rho}(A) := \rho A$, and $R_{\rho}(A) := A\rho$.  Since $L_{\rho}$
 and $R_{\rho}$ commute we may define $c(L_{\rho},R_{\rho})$.  Now we can
 state the fundamental theorems about monotone metrics.  In what
 follows uniqueness and classification are stated up to scalars (see Petz (1996)).

 \begin{Thm} (Chentsov 1982) 
     There exists a unique monotone metric on ${\cal P}_n^1$ given by the
     Fisher information.
 \end{Thm}

 \begin{Thm} (Petz 1996) 
     There exists a bijective correspondence between monotone metrics
     on ${\cal D}^1_n$ and symmetric operator monotone
     functions.  This correspondence is given by the formula
     $$
     g_{f} (A,B) :=g_{f,\rho} (A,B) :={\rm Tr}(A\cdot c_f(L_\rho,R_\rho)(B)).
     $$
\end{Thm}

Because of these two theorems we shall use the terms ``Monotone
Metrics" and ``Quantum Fisher Informations" with the same meaning.

Note that usually monotone metrics are normalized so that if
$[A,\rho]=0$ then $g_{f,\rho} (A,A)={\rm Tr}({\rho}^{-1}A^2)$, that is
equivalent to ask $f(1)=1$.

Examples of monotone metrics are given by the following list (see
Hasegawa and Petz (1997), Gibilisco and Isola (2004)).

Let
$$
f_\beta(x):= \beta (1- \beta) \frac{(x-1)^2}{(x^{\beta}-1)
(x^{1-\beta}-1)}\qquad \qquad \beta \in [-1,\frac{1}{2}] \backslash
\{0\} .
$$
$$
f_0(x):=\frac{x-1}{\log x}
$$
$$
h_{\gamma}(x) := \left( \frac{1+x^{\gamma}}{2}
\right)^{\frac{1}{\gamma}} \qquad \qquad \gamma \in [\frac{1}{2},1]
$$
Note that $f_0=\lim_{\beta \to 0}f_{\beta}$.

The $RLD$-metric is the QFI associated to $f_{-1}$.

The $BKM$-metric is the QFI associated to $f_0$.

The $WY$-metric is the QFI associated to $f_{\frac{1}{2}}=h_{\frac{1}{2}}$.

The $SLD$-metric (or Bures-Uhlmann metric) is the QFI associated to $h_1$.

The two parametric families $f_{\beta}, h_{\gamma}$ give us a
continuum of operator monotone functions from the smallest
$f_{-1}(x)=\frac{2x}{x+1}$ to the greatest $h_{1}=\frac{1+x}{2}$.

For a symmetric operator monotone function $\lim_{x \to +\infty}
\frac{f(x)}{x}=f(0)$.  Note that
$$
f_{\beta}(0)=0 \qquad \quad \beta \in [-1,0]
$$
$$
f_{\beta}(0)=\beta(1-\beta) \not=0 \qquad \qquad \beta \in (0,
\frac{1}{2}]
$$
$$
h_{\gamma}(0)=\left( \frac{1}{2} \right)^{\frac{1}{\gamma}} \not=0
\qquad \qquad \gamma \in [\frac{1}{2},1].
$$
The condition $f(0)\not=0$ is relevant because it is a necessary and
sufficient condition for the existence of the so-called radial
extension of a monotone metric to pure states (see Petz and Sudar
(1996)).

\section{Curvature for a Riemannian metric}
 
Let $V$ be a finite dimensional real vector space with a scalar
product $g(\cdot,\cdot)$.  We define, for $v,w \in V$,
$$
\hbox{Area}_g(v,w):=\sqrt{g(v,v) \cdot g(w,w)- \vert g(v,w) \vert^2}.
$$
In the euclidean plane $\hbox{Area}_g(v,w)$ is the area of the
parallelogramme spanned by $v$ and $w$.

For a linear connection $\nabla$ on a manifold $\cal M$ the curvature
is defined as (see Kobayashi and Nomizu (1963) pag.  133)
$$
R(X,Y)Z:=[\nabla_{X},\nabla_{Y}]Z-\nabla_{[X,Y]}Z.
$$
Suppose that $g(\cdot,\cdot)$ is a Riemannian metric on $\cal M$ and
$\nabla$ is the associated Levi-Civita connection.  The Riemannian
curvature tensor is defined as (see Kobayashi and Nomizu (1963) pag. 
201)
$$
R(X,Y,Z,W):=g(R(Z,W)Y,X)
$$
where $X,Y,Z,W$ are vector fields.

Now let $\rho \in {\cal M}$ and suppose that we have a 2-plane $\sigma
\in T_{\rho}{\cal M}$.  Then $\sigma$ determines a 2-dimensional
embedded surface ${\cal N}:={\rm exp}_{{\rho}}(B_{{\eta}}( 0_{\rho})
\cap \sigma)$ formed by the geodesic segments of length $<{\eta}$
which start tangentially to $\sigma$.  If $K(\sigma)$ denotes the
Gaussian curvature of $\cal N$ one has the following

\begin{Prop}\label{sectional}(see Klingenberg (1982) p.99) If
$A,B$ is a basis for the plane $\sigma$ then
$$
K(A,B):=K(\sigma)=\frac{R(A,B,A,B)}{g(A,A)g(B,B)-\vert g(A,B) \vert^2}
= \frac{g(R(A,B)B,A)} { {\rm Area}_g (A,B)^2} .
$$
\end{Prop}

When we want to emphasize the dependence of $R$ and $K$ from the
Riemannian metric $g$ we write $R_g$ and $K_g$.

\section{A geometric look at the inequality}

 Define $A_{\rho}:=i[\rho,A]$.  Since $A_{\rho}$ is traceless and
 selfadjoint, then $A_{\rho} \in T_\rho {\cal D}^1_n$.

 \begin{Prop} \label{disopra}
     For the QFI associated to $f_{\beta}$ one has
     $$
     g_{\beta}(A_{\rho},B_{\rho}):=g_{f_{\beta}}(A_{\rho},B_{\rho})=-\frac{1}{\beta
     (1- \beta)}{\rm Tr}([\rho^{\beta},A]\cdot[\rho^{1-\beta},B])
     \qquad \qquad \beta \in [-1,\frac{1}{2}] \backslash \{0\} \, .
     $$
\end{Prop}

One can find a proof in Hasegawa and Petz (1997), Gibilisco and Isola
(2004).  Because of the above proposition $g_{\beta}$ is known as the
WYD($\beta$) monotone metric.

If $f$ is an operator monotone function we denote by $R_{f}$ the
Riemannian curvature tensor, $K_{f}$ the sectional curvature and
Area$_{f}$ the area functional associated to the monotone metric
$g_{f}$.  Theorem \ref{K} takes the form

\begin{Thm}\label{MR}
    $\frac{({\beta}(1-{\beta}))^2}{4 } \left( {\rm
    Area}_{f_{\beta}}(A_{\rho},B_{\rho}) \right)^2 \hbox{ is an UPF
    for any } \, \beta \in (0,\frac{1}{2}]$.
\end{Thm}
\begin{proof}
One has from Theorem \ref{K}, Lemma \ref{Corr} and Proposition
\ref{disopra}
$$
{\cal S}_{\rho}(A,B) = {\rm Var}_{\rho}(A)\cdot{\rm
Var}_{\rho}(B)-|{\rm Re}\{{\rm Cov}_{\rho}(A,B) \}|^2
$$
$$
\geq I_{\rho,\beta}(A)I_{\rho,\beta}(B)-|{{\mathop{\rm Re}\nolimits}
\left\{ {\hbox{Corr}_{\rho,\beta} \left( {A,B} \right)} \right\}}|^2
$$
$$
=\left( -\frac{1}{2}{\rm Tr}([\rho^{\beta},A]\cdot[\rho^{1-\beta},A])
\right) \cdot \left( -\frac{1}{2}{\rm
Tr}([\rho^{\beta},B]\cdot[\rho^{1-\beta},B])\right)- \frac{1}{4}|{\rm
Tr}([\rho^{\beta},A]\cdot[\rho^{1-\beta},B])|^2=
$$
$$
=\frac{(\beta(1-\beta))^2}{4 }(g_{\beta}(A_{\rho},A_{\rho}) \cdot
g_{\beta}(B_{\rho},B_{\rho})- \vert
g_{\beta}(A_{\rho},B_{\rho})\vert^2)=\frac{(\beta(1-\beta))^2}{4}
\left( \hbox{Area}_{f_{\beta}}(A_{\rho},B_{\rho})\right)^2 \, .
$$
\end{proof}

Note that, if $\beta=\frac{1}{2}$, then
$K_{\frac{1}{2}}(\sigma)=costant=\frac{1}{4}$ (see Gibilisco and Isola
(2003)), so the inequality of Theorem \ref{LZ} takes the form
$$
{\cal S}_{\rho}(A,B) \geq \frac{1}{16} R_{ f_{ \frac{1}{2} }}
(A_{\rho},B_{\rho},A_{\rho},B_{\rho}) \, .
$$
In general from bounds on sectional curvature $K_{\beta}(\sigma)$ one
would be able to deduce inequalities of the same type for the Riemann
curvature tensor (see Gibilisco and Isola (2005) for ideas about this
kind of bounds).

At this point one may naturally ask: is the inequality
$$
{\cal S}_{\rho}(A,B) \geq \frac{(\beta(1-\beta))^2}{4} \left(
\hbox{Area}_{f_{\beta}}(A_{\rho},B_{\rho}) \right)^2
$$
true for other quantum Fisher informations of the WYD class?  For
example the $RLD$ metric corresponds to $\beta=-1$.  We shall see in
the next section that the answer is negative.

\section{A counterexample} \label{counterexample}

Theorem \ref{K} can be written as

\begin{Thm} \label{order2}
For any two self-adjoint operators $A$ and $B$, any density operator
$\rho$ and any $0 < \beta \leq \frac{1}{2} $, we have
$$
{\rm Var}_\rho \left( A \right){\rm Var}_\rho \left( B \right) -
\left| {{\mathop{\rm Re}\nolimits} \left\{ {{\rm Cov}_\rho \left(
{A,B} \right)} \right\}} \right|^2 \ge I_{\rho,\beta} \left( A
\right)I_{\rho,\beta} \left( B \right) - \left| {{\mathop{\rm
Re}\nolimits} \left\{ {{\rm Corr}_{\rho,\beta} \left( {A,B} \right)}
\right\}} \right|^2.
$$
\end{Thm}

\begin{proof}
We report here the proof of Yanagi {\sl et al.} (2005) because it is
needed in the sequel.

Let $\left\{\varphi_i\right\}$ be a complete orthonormal base composed
of eigenvectors of $\rho$, and $\{ {\lambda}_i \}$ the corresponding
eigenvalues.

Set $a_{ij} \equiv \langle {A_0} {\varphi}_i |{\varphi}_j \rangle $
and $ b_{ij} \equiv \langle B_0 \varphi_i | {\varphi_j } \rangle $.

Then we calculate
$$
\Var_{\rho}(A) = {\rm Tr} ( \rho A_0^2) = \frac{1}{2}\sum_{i,j} (
\lambda _i + \lambda_j ) a_{ij} a_{ji}
$$
$$
\Var_{\rho} ( B ) = {\rm Tr} (\rho B_0^2 ) = \frac{1}{2}\sum_{i,j}
(\lambda _i + \lambda _j )b_{ij} b_{ji}
$$
$$
{\rm Re}\{ \Cov_{\rho}(A,B) \} = {\rm Re} \{{\rm Tr} (\rho A_0 B_0)
\}= \frac{1}{2} \sum_{i,j}({\lambda}_i+{\lambda}_j) {\rm Re}
\{a_{ij}b_{ji} \}
$$
$$
I_{\rho,\beta} (A) = \Var_{\rho} ( A ) - {\rm Tr} ({\rho}^{\beta} A_0
{\rho}^{1-\beta} A_0) = \frac{1}{2}\sum_{i,j} ( \lambda _i + \lambda_j
) a_{ij} a_{ji} -\sum_{i,j} \lambda _i^\beta \lambda _j^{1 - \beta }
a_{ij}a_{ji}
$$
$$
I_{\rho,\beta} ( B ) =\frac{1}{2}\sum_{i,j} (\lambda _i + \lambda _j
)b_{ij} b_{ji} - \sum_{i,j} \lambda _i^\beta \lambda _j^{1 - \beta}
b_{ij} b_{ji}
$$
$$
{\rm Re} \{ {\rm Corr}_{\rho,\beta} ( {A,B} ) \} = {\rm Re} \{ {\rm
Cov}_{\rho} ( {A,B} ) \} - {\rm Re} \{ {\rm Tr}({\rho}^{\beta} A_0
{\rho}^{1-{\beta}} B_0) \}=
$$
$$
= \frac{1}{2} \sum_{i,j}({\lambda}_i+{\lambda}_j) {\rm Re} \{ a_{ij}
b_{ji} \} - \sum_{i,j} {\lambda}_i^\beta {\lambda}_j^{1 - \beta } {\rm
Re} \{ {a_{ij} b_{ji} } \} .
$$
Set
$$
\xi := \hbox{Var}_\rho \left( A \right)\hbox{Var}_\rho \left( B
\right) - I_{\rho,\beta} \left( A \right)I_{\rho,\beta} \left( B
\right)=
$$
$$
=\frac{1}{2}\sum_{i,j,k,l} \left\{ ( \lambda _i + \lambda _j )\lambda
_k^\beta \lambda _l^{1 - \beta } + ( \lambda _k + \lambda _l )\lambda
_i^\beta \lambda _j^{1 - \beta } - 2\lambda _i^\beta \lambda _j^{1 -
\beta } \lambda _k^\beta \lambda _l^{1 - \beta } \right\} a_{ij}
a_{ji} b_{kl} b_{lk}
$$
$$
=\frac{1}{4}\sum_{i,j,k,l} \left\{ ( \lambda _i + \lambda _j )\lambda
_k^\beta \lambda _l^{1 - \beta } + ( \lambda _k + \lambda _l )\lambda
_i^\beta \lambda _j^{1 - \beta } - 2\lambda _i^\beta \lambda _j^{1 -
\beta } \lambda _k^\beta \lambda _l^{1 - \beta } \right\} \{ a_{ij}
a_{ji} b_{kl} b_{lk}+ a_{kl}a_{lk}b_{ij}b_{ji} \}
$$
$$
\eta := \left| {{\mathop{\rm Re}\nolimits} \left\{ {\hbox{Cov}_\rho
\left( {A,B} \right)} \right\}} \right|^2 - \left| {{\mathop{\rm
Re}\nolimits} \left\{ {\hbox{Corr}_{\rho,\beta} \left( {A,B} \right)}
\right\}} \right|^2=
$$
$$
= \frac{1}{2}\sum_{i,j,k,l} \left\{ ( \lambda _i + \lambda _j )\lambda
_k^\beta \lambda _l^{1 - \beta } + ( \lambda _k + \lambda _l )\lambda
_i^\beta \lambda _j^{1 - \beta } - 2\lambda _i^\beta \lambda _j^{1 -
\beta } \lambda _k^\beta \lambda _l^{1 - \beta } \right\} {\rm Re} \{
a_{ij} b_{ji} \} {\rm Re} \{ a_{kl} b_{lk} \}.
$$
In order to prove the theorem it is enough to show $\xi - \eta \geq 0$. Indeed
$$
\xi - \eta =\frac{1}{4}\sum_{i,j,k,l} \left\{ ( \lambda _i + \lambda
_j )\lambda _k^\beta \lambda _l^{1 - \beta } + ( \lambda _k + \lambda
_l )\lambda _i^\beta \lambda _j^{1 - \beta } - 2\lambda _i^\beta
\lambda _j^{1 - \beta } \lambda _k^\beta \lambda _l^{1 - \beta }
\right\} \cdot
$$
$$
\cdot \left\{ |a_{ij} |^2 |b_{kl}|^2 + |a_{kl} |^2 |b_{ij}|^2 - 2{\rm
Re} \{ a_{ij} b_{ji} \} {\rm Re} \{ a_{kl} b_{lk} \} \right\}.
$$
Since
$$
\left( {\lambda _i + \lambda _j } \right)\lambda _k^\beta \lambda
_l^{1 - \beta } + \left( {\lambda _k + \lambda _l } \right)\lambda
_i^\beta \lambda _j^{1 - \beta } - 2\lambda _i^\beta \lambda _j^{1 -
\beta } \lambda _k^\beta \lambda _l^{1 - \beta } \\
=
$$
$$
= \left( \lambda _i + \lambda _j - \lambda _i^\beta \lambda _j^{1 -
\beta } \right)\lambda_k^\beta \lambda _l^{1 - \beta } + \left(
\lambda _k + \lambda _l -\lambda _k^\beta \lambda _l^{1 - \beta }
\right)\lambda _i^\beta \lambda _j^{1 - \beta } \ge 0 ,
$$
$$
|a_{ij} |^2 |b_{kl}|^2 + |a_{kl} |^2 |b_{ij}|^2 \ge 2\left| {a_{ij}
b_{ji} } \right|\left| {a_{kl} b_{lk} } \right| \ge 2\left|
{{\mathop{\rm Re}\nolimits} \left\{ {a_{ij} b_{ji} }
\right\}{\mathop{\rm Re}\nolimits} \left\{ {a_{kl} b_{lk} } \right\}}
\right| ,
$$
we get the thesis.
\end{proof}
	 
 \begin{Prop} \label{CE}
     For any $\beta \in [-1,0)$ there are a state $\rho$ and
     self-adjoint operators $A$ and $B$ s.t.
     $$
     {\rm Var}_\rho \left( A \right){\rm Var}_\rho \left( B \right) -
     \left| {{\mathop{\rm Re}\nolimits} \left\{ {{\rm Cov}_\rho \left(
     {A,B} \right)} \right\}} \right|^2 < I_{\rho,\beta} \left( A
     \right)I_{\rho,\beta} \left( B \right) - \left| {{\mathop{\rm
     Re}\nolimits} \left\{ {{\rm Corr}_{\rho,\beta} \left( {A,B}
     \right)} \right\}} \right|^2.
     $$
 \end{Prop}
\begin{proof}
Let $t\in(0,\frac{1}{2})$ and
$$
\rho = 
\left(\begin{array}{ccc}
    t & 0 & 0  \\
    0 & 1-2t & 0  \\
    0 & 0 & t
\end{array}\right), \quad
A=
\left(\begin{array}{ccc}
    0 & 1 & 0  \\
    1 & 0 & 0  \\
    0 & 0 & 0
\end{array}\right), \quad
B=
\left(\begin{array}{ccc}
    0 & 0 & 0  \\
    0 & 0 & 1  \\
    0 & 1 & 0
\end{array}\right).
$$	
Then, using the calculations performed for the proof of the previous theorem, we have
$$
\xi - \eta= \hbox{Var}_\rho(A) \hbox{Var}_\rho(B) - \left| {\rm Re}
\{\hbox{Cov}_\rho(A,B)\} \right|^2 - I_{\rho,\beta}(A)
I_{\rho,\beta}(B) - \left| {\rm Re} \{\hbox{Corr}_{\rho,\beta}(A,B)\}
\right|^2=
$$
$$
= \frac{1}{2} \sum\limits_{i,j,k,l} \left\{ (\lambda_i +
\lambda_j)\lambda_k^{\beta} \lambda_\ell^{1-{\beta}} + (\lambda_k +
\lambda_\ell)\lambda_i^{\beta} \lambda_j^{1-{\beta}} - 2\lambda_i^{\beta}
\lambda_j^{1-{\beta}} \lambda_k^{\beta} \lambda_\ell^{1-{\beta}}
\right\}\cdot
$$
$$
\qquad\cdot\left\{ a_{ij} a_{ji} b_{kl} b_{lk} - {\rm Re} (a_{ij}
b_{ji}) {\rm Re}(a_{kl} b_{lk}) \right\}=
$$
$$
= \frac{1}{2} (2\lambda_{1} + 2\lambda_{2} - \lambda_{1}^{{\beta}}
\lambda_{2}^{1-{\beta}} - \lambda_{2}^{{\beta}}
\lambda_{1}^{1-{\beta}}) (\lambda_{2}^{{\beta}}
\lambda_{3}^{1-{\beta}} + \lambda_{3}^{{\beta}}
\lambda_{2}^{1-{\beta}})+
$$
$$
+ \frac{1}{2} (2\lambda_{2} +2\lambda_{3} -\lambda_{2}^{{\beta}}
\lambda_{3}^{1-{\beta}} -\lambda_{3}^{{\beta}}
\lambda_{2}^{1-{\beta}}) (\lambda_{1}^{{\beta}}
\lambda_{2}^{1-{\beta}} + \lambda_{2}^{{\beta}}
\lambda_{1}^{1-{\beta}})
$$
$$
= \left\{ 2(1-t) -t^{{\beta}}(1-2t)^{1-{\beta}}
-(1-2t)^{{\beta}}t^{1-{\beta}}\right\}
\left\{t^{{\beta}}(1-2t)^{1-{\beta}}
+(1-2t)^{{\beta}}t^{1-{\beta}}\right\}.
$$

Let ${\beta}\in[-1,0)$.  Since $t^{{\beta}}(1-2t)^{1-{\beta}}\to
\infty$ if $t\to0^{+}$, there exists a $t_0=t_0(\beta) \in(0,1)$ for
which $\xi - \eta<0$.

This ends the proof.
\end{proof}

\begin{rem}
For $\beta \in [-1,0)$ the inequality
$$
\hbox{Var}_\rho \left( A \right)\hbox{Var}_\rho \left( B \right) -
\left| {{\mathop{\rm Re}\nolimits} \left\{ {\hbox{Cov}_\rho \left(
{A,B} \right)} \right\}} \right|^2 < I_{\rho,\beta} \left( A
\right)I_{\rho,\beta} \left( B \right) - \left| {{\mathop{\rm
Re}\nolimits} \left\{ {\hbox{Corr}_{\rho,\beta} \left( {A,B} \right)}
\right\}} \right|^2
$$
is not true in general as one can see by choosing
$$
t\in(0,1), \quad \rho = 
\left(\begin{array}{cc}
    t & 0  \\
    0 & 1-t
\end{array}\right), \quad A =
\left(\begin{array}{cc}
    1 & 0  \\
    0 & 0
\end{array}\right), \quad B = 
\left(\begin{array}{cc}
    0 & 0  \\
    0 & 1
\end{array}\right).
$$
\end{rem}

\section{Open problems}

{\bf Problem 1}

The counterexample of Proposition \ref{CE} seems a definitive result
that forbids further generalizations of the inequality of Theorem
\ref{K}.

Maybe one should seek a different generalization of Theorem \ref{K}. 
Since
$$
\beta \in [-1,0] \Longrightarrow f_{\beta}(0)=0 \qquad \& \qquad \beta
\in (0,\frac{1}{2}]\Longrightarrow f_{\beta}(0)=\beta(1-\beta)
$$
one can state Theorem \ref{K} (that is Theorem \ref{order2} or Theorem
\ref{MR}) in a different way

\begin{Thm}
$$
{\cal S}_{\rho}(A,B) \geq \frac{f_{\beta}(0)^2}{4 } \left( {\rm
Area}_{f_{\beta}}(i[A,\rho],i[B,\rho]) \right)^2 \qquad \qquad \forall
\beta \in [-1,\frac{1}{2}]
$$
\end{Thm}

Question: characterize the family of operator monotone functions $f$
for which is true the inequality

$$
{\cal S}_{\rho}(A,B) \geq \frac{f(0)^2}{4 } \left( {\rm
Area}_{f}(i[A,\rho],i[B,\rho]) \right)^2.
$$

Of course the above inequality is trivially true when $f(0)=0$ while
it is a non-trivial inequality for those operator monotone functions
such that $f(0)>0$.  Note that the question is non-trivial, for
example, for the $SLD$ metric for which $h_1(0)=\frac{1}{2}$.

{\bf Problem 2}

For $f$ operator monotone define
$$
G(f):=\frac{f(0)^2}{4 } \left( {\rm Area}_{f}(i[A,\rho],i[B,\rho])
\right)^2.
$$
In Kosaki (2005) the proof of Theorem \ref{order2} (see p.640) is
obtained by the following result
$$
f_{\beta} \leq f_{\tilde{\beta}} \Longrightarrow G(f_{\beta}) \leq
G(f_{\tilde{\beta}}) \qquad \qquad \beta \in (0,\frac{1}{2}]
$$

Is this inequality true for other families of operator monotone
functions?

{\bf Problem 3}.

 The following question has been posed at p.642 in Kosaki (2005). 
 Covariance and $WYD$ information make perfect sense in infinite
 dimension (see Connes and Stormer (1978), Kosaki (1982)), namely in a
 general von Neumann algebra setting.  Is the inequality of Theorem
 \ref{order2} still true in this general setting?

\begin{ack}
    The present results have been presented at the second International
Conference ``Information Geometry and its Applications" held at the
University of Tokio, December 12-16, 2006.  It is a pleasure to thank
all the organizers.
\end{ack}

 \section{References}

 {H.~Araki and M.~M.~Yanase} (1960). 
{Measurement of quantum mechanical operators},
\textit{ Physical Review}, \textbf{(2)120},
{622}--{626}.

{A.~Connes and E.~Stormer} (1978). {Homogeneity of the state space of factors of type $III_1$},
\textit{Journal of  Functional Analysis}, \textbf{28}, {187}--{196}.

{P.~Gibilisco and T.~Isola} (2003).  {Wigner-Yanase information on
quantum state space: the geometric approach}, \textit{Journal of
Mathematical Physics}, \textbf{44(9)}, {3752}--{3762}.

{P.~Gibilisco and T.~Isola} (2004). {On the characterisation of paired monotone metrics},
\textit{Annals of the Institute of  Statistical Mathematics}, \textbf{56(2)}, {369}--{381}.

{P.~Gibilisco and T.~Isola} (2005).  {On the monotonicity of scalar
curvature in classical and quantum information geometry},
\textit{Journal of Mathematical Physics}, \textbf{46(2)},
{023501}--{14}.

{P.~Gibilisco and T.~Isola} (2006). {Some open problems in Information Geometry}, To appear in
\textit{``Proceedings 26th Conference: QP and IDA" - Levico (Trento), February 20-26, 2005}.

{H.~Hasegawa and D.~Petz} (1997).  {Non-commutative extension of the
information geometry II}, \textit{Quantum Communications, Computing
and Measurement} (eds.\ {O. Hirota et al.}), {109}--{118}, {Plenum},
{New York}.

{W.~Klingenberg} (1982). 
\textit{Riemannian Geometry},
{Walter de Gruyter \& Co.}, {Berlin}.

{S.~Kobayashi and K. ~Nomizu} (1963). 
\textit{Foundations of differential geometry, Vol. I.},
{John Wiley \& Sons}, {New York-London}.

{H.~Kosaki} (1982). {Interpolation theory and the Wigner-Yanase-Dyson-Lieb concavity},
\textit{Communications in  Mathematical Physics}, \textbf{87}, {315}--{329}.

{H.~Kosaki} (2005). {Matrix trace inequalities related to uncertainty principle},
\textit{International Journal of Mathematics}, \textbf{6}, {629}--{645}.

{S.~Luo and Q. Zhang} (2004). {On skew information},
\textit{IEEE Transactions on Information Theory}, \textbf{50(8)}, {1778}--{1782}.

{S.~Luo and Z. Zhang} (2004).  {An informational characterization of
Schr{\"o}dinger's uncertainty relations}, \textit{Journal of
Statistical Physics}, \textbf{114(5-6)}, {1557}--{1576}.

{M.~Ozawa} (2002). {Conservation laws, uncertainty relations and quantum limits of measurement},
\textit{Physical Review Letters}, \textbf{88(5)}, {050402}--{4}.

{D.~Petz} (1996). {Monotone metrics on matrix spaces},
\textit{Linear Algebra and Applications}, \textbf{244}, {81}--{96}.

{D.~Petz and C. ~ Sudar} (1996). {Geometries of quantum states},
\textit{Journal of Mathematical Physics}, \textbf{37(6)}, {2662}--{2673}.

{E. ~Schr{\"o}dinger (1930).  {About Heisenberg uncertainty relation
(original annotation by A. Angelow and M.C. Batoni)},
\textit{Bulgarian Journal of Physics}, \textbf{26 (5-6)}, {193}--{203}
(2000), 1999.  Translation of \textit{Proceedings Prussian Academy of
Sciences, Physical and Mathematical Section} \textbf{19} (1930),
{296}--{303}.

{E. ~ P.~ Wigner} (1952). {Die Messung quantenmechanischer Operatoren},
\textit{Zeitschrift fur Physics}, \textbf{133}, {101}--{108}.

{K.~Yanagi, S. Furuichi and K. Kuriyama} (2005).  {A generalized skew
information and uncertainty relation}, \textit{IEEE Transactions on
Information Theory}, \textbf{51(12)}, {4401}--{4404}.

\end{document}